\documentclass[prd,aps,twocolumn,amsmath,amssymb,nofootinbib,preprintnumbers]
{revtex4}

\begin{document}
\newcommand{\drawsquare}[2]{\hbox{%
\rule{#2pt}{#1pt}\hskip-#2pt
\rule{#1pt}{#2pt}\hskip-#1pt
\rule[#1pt]{#1pt}{#2pt}}\rule[#1pt]{#2pt}{#2pt}\hskip-#2pt
\rule{#2pt}{#1pt}}

\newcommand{\Yfund}{\raisebox{-.5pt}{\drawsquare{6.5}{0.4}}}
\newcommand{\Yasymm}{\raisebox{-3.5pt}{\drawsquare{6.5}{0.4}}\hskip-6.9pt%
                     \raisebox{3pt}{\drawsquare{6.5}{0.4}}%
                    }
\newcommand{\Ysymm}{\Yfund\hskip-0.4pt%
                    \Yfund}
\def\symm{\Ysymm}
\def\bsymm{\overline{\Ysymm}}
\def\ls{\mathrel{\lower4pt\vbox{\lineskip=0pt\baselineskip=0pt
           \hbox{$<$}\hbox{$\sim$}}}}
\def\gs{\mathrel{\lower4pt\vbox{\lineskip=0pt\baselineskip=0pt
           \hbox{$>$}\hbox{$\sim$}}}}
\def\drawbox#1#2{\hrule height#2pt
        \hbox{\vrule width#2pt height#1pt \kern#1pt
              \vrule width#2pt}
              \hrule height#2pt}

\def\Fund#1#2{\vcenter{\vbox{\drawbox{#1}{#2}}}}
\def\Asym#1#2{\vcenter{\vbox{\drawbox{#1}{#2}
              \kern-#2pt       
              \drawbox{#1}{#2}}}}
\def\sym#1#2{\vcenter{\hbox{ \drawbox{#1}{#2} \drawbox{#1}{#2}    }}}
\def\fund{\Fund{6.4}{0.3}}
\def\asymm{\Asym{6.4}{0.3}}
\def\bfund{\overline{\fund}}
\def\basymm{\overline{\asymm}}


\newcommand{\beq}{\begin{equation}}
\newcommand{\eeq}{\end{equation}}
\def\ls{\mathrel{\lower4pt\vbox{\lineskip=0pt\baselineskip=0pt
           \hbox{$<$}\hbox{$\sim$}}}}
\def\gs{\mathrel{\lower4pt\vbox{\lineskip=0pt\baselineskip=0pt
\def\lsim{\mathrel{\lower4pt\vbox{\lineskip=0pt\baselineskip=0pt
           \hbox{$<$}\hbox{$\sim$}}}}
\def\gsim{\mathrel{\lower4pt\vbox{\lineskip=0pt\baselineskip=0pt
           \hbox{$>$}\hbox{$\sim$}}}}           \hbox{$>$}\hbox{$\sim$}}}}

\title{Longevity of Supersymmetric Flat Directions}

\author{Rouzbeh Allahverdi$^{1,2}$}
\author{Anupam Mazumdar$^{3}$}
\affiliation{$^{1}$~Perimeter Institute for Theoretical Physics, Waterloo, ON,
N2L 2Y5, Canada. \\ $^{2}$~Department of Physics and Astronomy,
McMaster University, Hamilton, ON, L8S 4M1, Canada. \\ $^{3}$~NORDITA,
Blegdamsvej-17, Copenhagen-2100, Denmark.}

\begin{abstract}
We examine the fate of supersymmetric flat directions. We argue that
the non-perturbative decay of the flat direction via preheating is
an unlikely event. In order to address this issue, first we identify
the physical degrees of freedom and their masses in presence of a
large flat direction VEV (Vacuum Expectation Value). We explicitly
show that the (complex) flat direction and its fermionic partner are
the only light {\it physical} fields in the spectrum. If the flat
direction VEV is much larger than the weak scale, and it has a
rotational motion, there will be no resonant particle production at
all. The case of multiple flat directions is more involved. We
illustrate that in many cases of physical interest, the situation
becomes effectively the same as that of a single flat direction, or
collection of independent single directions. In such cases
preheating is not relevant. In an absence of a fast non-perturbative
decay, the flat direction survives long enough to affect
thermalization in supersymmetric models as described in
hep-ph/0505050 and hep-ph/0512227. It can also ``terminate'' an
early stage of non-perturbative inflaton decay as discussed in
hep-ph/0603244.
\end{abstract}

\maketitle

The Minimal Supersymmetric Standard Model (MSSM) has a large number of
flat directions, along which the scalar potential vanishes. The flat
directions are parameterized by {\it gauge-invariant}
monomials~\cite{DRT,GKM}, for a review see \cite{MSSM-REV}.  Recently
it has been pointed out that these flat directions can play an
important role in thermal history of the Universe within supersymmetry
(SUSY)~\cite{AVERDI2,AVERDI1}. We remind that the thermalization rate
depends on $2 \leftrightarrow 2$ and $2 \rightarrow 3$ scatterings of
the inflaton decay products. The former lead to kinetic equilibrium
but do not change the total number of particles.  However, after the
last stage of (perturbative) inflaton decay, the reheat plasma is far
from chemical equilibrium as the number of particles is much less than
that in full thermal equilibrium~\cite{AVERDI2}. As pointed out
in~\cite{ds}, $2 \rightarrow 3$ scatterings with gauge boson exchange
in the $t$-channel play the main role in establishing chemical
equilibrium. For massless gauge bosons, they are infrared
divergent. In a non-supersymmetric set up the plasma mass of gauge
bosons provides a natural infrared cut-off which regulates these
processes(for details see~\cite{thermalization}). Since right after
the inflaton decay the reheat plasma is dilute, the resulting plasma
mass is small. Thus $2 \rightarrow 3$ scatterings are enhanced and
additional particles are efficiently produced. In consequence, full
thermal equilibrium is established quickly in this case~\cite{ds}.

However the situation will be completely different within
supersymmetry because of the flat directions. During inflation these
flat directions develop large Vacuum Expectation Value (VEV), see for
details~\cite{MSSM-REV}. A large VEV induces large SUSY preserving
masses to gauge bosons and gauginos via the Higgs
mechanism~\cite{AVERDI2}~\footnote{There are ways to suppress the flat
direction VEV, if the MSSM flat direction is heavy during inflation
then the flat direction settles down to the minimum of its
potential. However this happens for a very special case of the
K\"ahler potential, see for details~\cite{MSSM-REV}.}.  The $2
\rightarrow 3$ scatterings mediated by the gauge bosons are therefore
strongly suppressed and, consequently, thermalization will be very
slow~\cite{AVERDI1,AVERDI2}. We also showed that for a Standard Model
(SM) {\it singlet inflaton} the inflaton decay is typically {\it
perturbative} within SUSY~\cite{AVERDI2,AVERDI3}~\footnote{In almost
all models of inflation, the inflaton is treated as a SM gauge
singlet. Very recently we constructed models of inflation which are
{\it gauge invariant} combinations of squarks and
sleptons~\cite{AEGM}, and with right-handed sneutrinos~\cite{AKM}. The
model predictions are trustable and robust (see for other gauge
invariant inflatons~\cite{JM}). We emphasize that this novelty is {\it
solely} due to the fact that the inflaton is charged under the MSSM
gauge group.}.

Delayed thermalization has important outcomes: the Universe undergoes
a quasi-thermal phase for a long period after the inflaton has
completely decayed, during which the reheat plasma is dilute and far
from chemical equilibrium~\cite{AVERDI1,AVERDI2}.  After the flat
direction starts oscillating, its VEV and its induced mass for gauge
bosons/gauginos is redshifted by the Hubble expansion. Gradually
thermalization rate increases and full thermal equilibrium is
established shortly after this rate becomes comparable to the
expansion rate of the Universe. A full thermalization typically
happens much later than the inflaton decay, leading to a {\it reheat
temperature} (much) lower than naively expected~\cite{AVERDI2}. The
exact value of the reheat temperature depends on the nature of the
flat direction and its initial VEV and, typically it is in the range
$T_{\rm R}\sim 10^{4}-10^{7}$~GeV~\cite{AVERDI2}.

The notable off-shoots of a delayed thermalization and a low reheat
temperature are~\footnote{A late thermalization holds the key to a
successful MSSM curvaton mechanism~\cite{AEJM}, see also~\cite{ENQ}.}:

\begin{itemize}

\item{Thermal gravitino production is suppressed. SUSY provides a
natural solution to the infamous thermal gravitino
problem~\cite{AVERDI1,AVERDI2}.  Note that the abundance of
non-thermal gravitinos is not affected~\cite{MAROTO,MAR}, but they do
not give rise to a large gravitino abundance any way.}

\item{Thermal leptogenesis within MSSM does not work in most of the
parameter space~\cite{AVERDI1,AVERDI2}.}

\end{itemize}

The above conclusions hold if the flat direction VEV remains large for
a sufficiently long time {\it after} the onset of its
oscillations. This is the case if the flat direction decays
perturbatively. Note that the flat direction VEV induces a large mass
to all fields which are coupled to it. Perturbative one-particle decay
to these fields is only possible when their mass, hence the flat
direction VEV, is less than the flat direction mass. If the initial
VEV is much larger than the weak scale, this happens much later than
the oscillations start.

Recently it has been pointed out in~\cite{OP} that, the flat direction
oscillations can decay very rapidly (i.e., within a few oscillations)
via preheating~\cite{BRANDENBERGER,LINDE}. This will have important
implications for cosmology. For example the Universe will quickly
thermalize even in the presence of a flat direction with a large
VEV. Here we examine this possibility in some detail and argue that
the decay of MSSM flat directions via preheating is unlikely in cases
of physical interest.

First of all, we identify the physical degrees of freedom and the
flat direction masses when a single flat direction has a non-zero
VEV. We show that the flat direction and its fermionic partner are
the only light fields in the relevant spectrum. As a result, the
flat direction motion, so long as it has a rotational component,
results in time variation of the mass eigenstates and eigenvalues of
{\it superheavy} fields. This, however, is not significant at all
and does not lead to resonant particle production~\cite{ACS,MP}.

Next, we consider the case with more than one flat directions. Two
or more flat directions which are simultaneously $D$- and $F$-flat
can develop large VEVs at the same time. Within MSSM the only
directions which are completely independent are the two represented
by the ${LLe}$ and ${udd}$ monomials. In this case both the flat
directions evolve independently and their dynamics is equivalent to
a single flat direction, and hence there will be no preheating. We
also emphasize that preheating will not be relevant when two flat
directions have very different VEVs. This happens in many cases
since the directions under consideration are lifted by different
$F$-terms which become important at different field values.

The second possibility is multiple flat directions associated with a
gauge-invariant polynomial (when all families of leptons and quark are
taken into account). This is a new case and very involved, even for a
simple example of ${H_u L}$ monomial, whose dynamics was first
described in Ref.~\cite{JKM}, but preheating effects were not
discussed. In a general case, however, the motion of the flat
direction VEV is not known {\it yet}. One cannot embark on the issue
of particle creation before knowing the details of the flat direction
motion for such a scenario. We will briefly sketch this scenario and
we think preheating is unimportant for practical purposes. Moreover,
due to $F$-term lifting, in many cases the flat manifold has complex
dimension {\it one} at very large field values. In such cases the
situation becomes similar to that of a single flat direction.

The central message is: {\it preheating of an MSSM flat direction is
unimportant for physically interesting cases, thus, the flat direction
lives long enough to affect reheating and thermalization}.

\section{Spontaneous symmetry breaking and physical degrees of freedom}

A crucial point is to identify the physical degrees of freedom and
their mass spectrum in presence of a non-zero flat direction VEV
~\footnote{The first version of Ref.~\cite{OP} did not {\it count}
correctly the light degrees of freedom. Then it was claimed that even
a single MSSM flat direction can decay non-perturbatively.  However,
as we argue below, this is not the case.}.

Let us consider the simplest flat direction, which includes only two
fields: ${H_u H_d}$. This is also familiar from the electroweak
symmetry breaking in MSSM. A clear and detailed discussion is given
in~\cite{HK}. Here we briefly recount the situation.

One can always rotate the field configuration to a basis where, up to
an overall phase, $H^1_{u} = H^2_{d} = 0$ and $H^2_{u} = H^1_{d} =
\phi_0/\sqrt{2}$.Here superscripts denote the weak isospin components
of the Higgs doublets. In this basis the complex scalar field is
defined by:
\beq \label{flat}
\varphi = {(H^2_{u} + H^1_{d}) \over \sqrt{2}}\,,
\eeq
represents a flat direction. Its VEV breaks the $SU(2)_W \times
U(1)_Y$ down to $U(1)_{\rm em}$~(in exactly the same fashion as in the
electroweak vacuum). The $W^{\pm}$ and $Z$ gauge bosons then obtain
masses $m_W,~m_Z \sim g \phi_0$ from their couplings to the Higgs
fields via covariant derivatives ($g$ denotes a general gauge
coupling).

We also have:
\beq \label{gold}
\chi_1 = {(H^2_{u} - H^1_{d}) \over \sqrt{2}}\,,~~
{\rm and}~~
\chi_2 = {(H^1_{u} + H^2_{d}) \over \sqrt{2}}\,.
\eeq
Then $\chi_2$ and $\chi_{1,~R}$~($R$ and $I$ denote the real and
imaginary parts of a complex scalar field respectively) acquire masses
equal to $m_W$ and $m_Z$, respectively, through the $D-$term part of
the scalar potential. Note that
\beq
\chi_3 = {(H^1_{u} - H^2_{d}) \over \sqrt{2}}\,,
\eeq
and $\chi_{1,~I}$ are the three Goldstone bosons, which are eaten by
the gauge fields via the Higgs mechanism. Therefore, out of $8$ real
degrees of freedom in the two Higgs doublets, there are only two light
{\it physical} fields: $\varphi_{R},~\varphi_{I}$. They are exactly
massless when SUSY is not broken (and there is no $\mu$ term either).

An important point is that the masses induced by the flat direction
VEV are SUSY conserving. One therefore finds the same mass spectrum in
the fermionic sector. More specifically, the Higgsino fields
${\widetilde H}^1_{u}$ and ${\widetilde H}^2_{d}$ are paired with the
Winos, while $({\widetilde H}^2_{u} - {\widetilde H}^1_{d})/\sqrt{2}$
is paired with the Zino to acquire masses equal to $m_W$ and $m_Z$,
respectively, through the gaugino-gauge-Higgsino interaction terms.
The fermionic partner of the flat direction $({\widetilde H}^2_{u} +
{\widetilde H}^1_{d})/\sqrt{2}$ remains massless (note that the photon
and photino are also massless, but not relevant for our discussion).

In reality, supersymmetry is broken and, $\varphi$ obtains a mass
$m_{\varphi} \sim {\cal O}({\rm TeV})$ from soft SUSY breaking term
(the same is true for the gauginos). However, for $g \varphi_0 \gg
{\cal O}({\rm TeV})$, which is the situation relevant to the early
Universe, the mass spectrum is hierarchical: $\chi_{1,~R}$,
$\chi_{2}$, and gauge fields (plus their fermionic partners) are
superheavy.

In a general case the total number of light scalars, $N_{light}$, is
given by:
\beq \label{lightnum}
N_{light} = N_{total} - (2 \times N_{broken}),
\eeq
where $N_{total}$ is the total number of scalar degrees of freedom,
and $N_{broken}$ is the number of spontaneously broken symmetries.
Note that the factor $2$ counts for the number of eaten Goldstone
bosons plus the number of degrees of freedom which have obtained large
masses equal to those of the gauge bosons. In the case of $H_u H_d$
direction, Eq.~(\ref{lightnum}) reads:~$2 = 8 - (2 \times 3)$.


\section{Particle production from a single flat direction}

A generic MSSM flat direction has $F$-term and $D$-term couplings to
other fields. The dynamics and non-perturbative particle creation for
a flat direction with $F$-term couplings are unimportant. This has
been clearly illustrated in Refs.~\cite{AC,MP}, here we will not
discuss this case. We will only concentrate on the $D$-terms.

For the ${H_u H_d}$ flat direction, supersymmetric $D$-term
contribution lead to:
\begin{eqnarray} \label{dterm}
& V_D = (a_1 g^2_1 + a_2 g^2_2) \left[\varphi_R \chi_{1,~R} + \varphi_I
\chi_{1,~I} \right]^2 + \, \nonumber \\
& a_2 g^2_2 \left[\varphi_R \chi_{2,~R} + \varphi_I \chi_{2,~I} \right]^2 + a_2
g^2_2 \left[\varphi_R \chi_{3,~R} + \varphi_I \chi_{3,~I} \right]^2\,,
\nonumber \\
\end{eqnarray}
with $a_1,~a_2\sim {\cal O}(10^{-1})$. Here $g_1$ and $g_2$ are gauge
couplings of $U(1)_Y$ and $SU(2)_W$.  For a general monomial the
$D$-term contribution will be similar when the scalars are decomposed
to the flat direction $\varphi$ and orthogonal fields $\chi$.

Let us now consider the case where the flat direction VEV is
oscillating. The oscillations start when the Hubble expansion rate
drops below the flat direction mass, i.e., when $H (t)\simeq
m_{\varphi}$. Note that the motion is due to soft supersymmetry
breaking terms and the $F$-terms~\cite{DRT}, as $V_D$ identically
vanishes for a flat direction. The soft mass and superpotential
contribution depends on the modulus of flat direction $\vert \varphi
\vert$. On the other hand, the $''A''$-terms result in a
phase-dependent contribution, which exerts a torque and generates a
rotation in the $\varphi$ plane. As a generic feature the flat
direction motion has a major rotational component~\cite{DRT}.

\subsection{Rotation on a circle}

If the real and imaginary parts of the flat direction oscillate with
the same amplitude, $\varphi_0$, but with a $\pi/2$ phase difference,
the trajectory will be a circle in the $\varphi$ plane: $\varphi
=\varphi_0 {\rm exp}(i m_{\varphi}t)$. This is not what exactly
happens in the early Universe, but as we just pointed out, this is a
very good approximation of a generic situation.

The first term on the right-hand side of Eq.~(\ref{dterm}), which
arises from the diagonal part of $SU(2)_W$ and $U(1)_Y$, gives rise
to~\footnote{Contributions from non-diagonal parts of $SU(2)_W$ lead
to a similar contribution with $\chi_1$ being replaced by
$\chi_{2},~\chi_{3}$.}:
\beq \label{hl}
V \sim (g^2_1 + g^2_2) ~ \varphi^2_0 ~ \left[{\rm \cos}(m_{\varphi}t)
\chi_{1,~R} + {\rm sin}(m_{\varphi}t) \chi_{1,~I} \right]^2 \,.
\eeq
At a first glance it seems that there are two scalar mass eigenstates:
\begin{eqnarray} \label{eigst}
\vec{\omega}_1 (t) = {\rm cos}(m_{\varphi}t) \chi_{1,~R} + {\rm
sin}(m_{\varphi}t)
\chi_{1,~I} \,, \nonumber \\
\vec{\omega}_2 (t) = {\rm cos}(m_{\varphi}t) \chi_{1,~I} - {\rm
sin}(m_{\varphi}t) \chi_{1,~R} \,,
\end{eqnarray}
where $\omega_1$ has a mass $\sim g \varphi_0$, hence superheavy,
and $\omega_2$ is light. Note that the mass eigenvalues are constant
in time (despite rotation of $\varphi$) but the mass eigenstates are
evolving.

In general one can picture an instantaneous mass eigenstate by a
vector, $\vec{\omega}(t)$, whose magnitude represents the
corresponding mass eigenvalue. In the case at hand the mass
eigenstates are two-dimensional vectors in the $\chi_{1}$ plane.  One
can then expect non-perturbative production of particles whenever the
adiabaticity is violated in the evolution of $\vec{\omega}$, i.e.,
such that
\beq \label{nonad}
\left\vert {d \vec{\omega}(t) \over dt}\right\vert
\gs {\vert \vec{\omega}(t) \vert}^2.
\eeq
It can be shown from Eq.~(\ref{eigst}) that this is the case at all
moments of time for the light mass eigenstate $\vec{\omega}_2(t)$.
One might then conclude that quanta of the light mode are copiously
produced via parametric resonance, thus, leading to a quick decay of
the flat
direction oscillations~\cite{OP}.\\

{\it However, a closer inspection shows that $\vec{\omega}_2 (t)$
actually is one of the three Goldstone bosons in the background of
rotating
$\varphi$, see Eq.~(\ref{gold}), and the subsequent discussions.}\\

Hence $\vec{\omega}_2(t)$ is not a {\it physical} degree of freedom
as it is eaten by the $Z$ gauge boson. Indeed, as we elaborated,
there are no {\it physical} light fields except the rotating flat
direction. Obviously the mass eigenstates and eigenvalues of the
flat direction are not affected by its rotation and therefore they
are {\it constant} in time, see Eq.~(\ref{flat}).

Further note that the rotation only results in a time variation in the
mass eigenstates of the heavy fields, and the corresponding
eigenvalues (which are constant in time) are much larger than the
frequency of rotation.

At the onset of the flat direction oscillations, we typically have,
$g \varphi_0 \geq 10^{8} m_{\varphi}$. This implies that the {\it
adiabaticity condition} always holds, see Eq.~(\ref{nonad}), and
hence there will be no resonant particle production~\footnote{Note
that for a circular rotation, the Lagrangian for the heavy fields is
the same as that in a static case up to corrections of order
$(m_{\varphi}/g \varphi_0)$, coming form time-derivatives of
$\varphi$.  Therefore, one can reliably use the same mass
eigenstates and eigenvalues as that of the static case.}.

In an absence of particle production, the flat direction VEV is only
subject to redshifting by the Hubble expansion. In a Universe
dominated by relativistic particles, not necessarily thermalized, we
have $\langle \varphi \rangle \propto H^{3/4}$. Therefore, the heavy
fields become continuously lighter.

The resonant particle production, due to time-variation of the mass
eigenstates, if at all becomes {\it important}, will only be relevant
when $g \langle \varphi\rangle \sim m_{\varphi}$. However, this will
happen much later than the onset of the flat direction oscillations,
indeed, we find:
\beq \label{preheat}
H_{preheat} \sim
\left({m_{\varphi} \over g \varphi_0} \right)^{4/3} m_{\varphi}
\leq 10^{-9} m_{\varphi},
\eeq
where $m_{\varphi}\sim {\cal O}(\rm TeV)$.  This implies that the flat
direction VEV survives a very long time, which is sufficient to affect
thermalization along the lines pursued in
Refs.~\cite{AVERDI1,AVERDI2}. In particular, Eq.~(\ref{preheat}) can
be translated into an upper bound on the reheat temperature: $T_{\rm
R} \sim (H_{preheat} M_{\rm P})^{1/2} \leq 10^6$ GeV. This is
compatible even with the most stringent limit on the reheat
temperature, which is obtained for unstable gravitinos with a dominant
hadronic decay mode~\cite{kkm}.

\subsection{A General Rotation}

For a general rotation we will have: $\varphi_R =\varphi_0~{\rm
\cos}(m_{\varphi}t)$ and $\varphi_I = a \varphi_0~{\rm
\cos}(m_{\varphi}t + \theta)$, where $a \leq 1$ is a positive number
and $\theta$ is an ${\cal O}(1)$ phase.

In this case, both the mass eigenstates and mass eigenvalues of the
heavy fields change in time. The mass eigenvalues, given by $\sim g
\vert \varphi \vert$, lie in the range $\sim [a g \varphi_0 , g
\varphi_0]$. Particle production will be insignificant so long as the
time evolution of the system is adiabatic. The adiabaticity condition
for mass eigenvalues is violated if:
\beq \label{nonad2}
g {d \vert \varphi \vert \over dt} \gs g^2 {\vert \varphi \vert}^2.
\eeq
This happens when $g \vert \varphi \vert \ls a (g \varphi_0
m_{\varphi})^{1/2}$, and provided that $a < (m_{\varphi}/g
\varphi_0)^{1/2}$~\cite{ACS,MP}.  For a typical initial condition,
where $g\varphi_0 \geq 10^8 m_{\varphi}$, this requires that, $a \leq
10^{-4}$. As mentioned earlier, in a generic situation we have $a \sim
{\cal O}(1)$~\cite{DRT}.

For such small values of $''a''$, the flat direction motion will be
effectively one-dimensional: $\varphi_R \approx \varphi_0 {\rm
cos}(m_{\varphi}t)$ and $\varphi_I \approx 0$. In this case, $D$-terms
result in an interaction term $g^2 \varphi^2_R \chi^2_R$, see
Eq.~(\ref{hl}), which leads to a fast decay of flat direction
oscillations via the standard picture of preheating~\cite{LINDE}.
However, the challenge is to seek, $a\leq 10^{-4}$, in a physical
models.

{\it We conclude that in general; the flat direction oscillations
survive a very long time. A fast decay can only happen in the
exceptional case when the flat direction motion is effectively
one-dimensional.}

\section{Two or more flat directions}

A necessary (but not sufficient) condition for a quick decay of an
MSSM flat direction via preheating is that; there must exist light
scalar degrees of freedom whose mass eigenstates and/or eigenvalues
are evolving in time in a non-adiabatic fashion. As we showed this
does not happen for a single flat direction. Here we examine the
situation where there are multiple flat directions. We discuss two
possible cases separately.


\subsection{Two directions represented by different monomials}

Two directions, which are simultaneously $D$- and $F$-flat, can
independently acquire large VEVs during inflation.  An an important
example, there exists two flat directions represented by the ${udd}$
and ${LLe}$ monomials, respectively. Note that ${u}$ and ${d}$ have no
Yukawa couplings to ${L}$ and ${e}$. This guarantees that they are
simultaneously $F$-flat. The $D-$flatness imposes that the
two ${d}$'s in ${udd}$, and the two ${L}$'s in ${LLe}$ are from
different families. Also $F$-flatness requires that ${e}$ belong to a
third lepton family. Without any loss of generality, let us consider
two flat directions, parameterized as:
\begin{eqnarray}
& u^1_1 = d^2_1 = d^3_2 = {\varphi_0 \over \sqrt{3}}, ~ ~ ({\rm up ~to ~a~
phase}) \,, \nonumber \\
& \, \nonumber \\
& L^1_1 = L^2_2 = e_3 = {\varphi^{\prime}_0 \over \sqrt{3}}, ~ ~ ({\rm up ~to~
 a ~phase}). \,
\end{eqnarray}
Here the upper indices indicate color (for ${u}$ and ${d}$) and weak
isospin (for ${L}$). The lower indices denote the family. Note that
${u}$ can be from any of the quark families, and we have chosen, $1$,
just as one possible case. Then the flat directions $\varphi$ and
$\varphi^{\prime}$ are given by:
\beq \label{flat1}
\varphi = {(u^1_1 + d^2_1 + d^3_2 ) \over \sqrt{3}} ~ ~ , ~ ~ \varphi^{\prime}
= {(L^1_1 + L^2_2 + e_3) \over \sqrt{3}},
\eeq
with respective masses:
\beq \label{mass1}
m^2_{\varphi} = {(m^2_{u_1} + m^2_{d_1} + m^2_{d_2}) \over 3} ~ ~ , ~ ~
m^2_{\varphi^{\prime}} = {(m^2_{L_1} + m^2_{L_2} + m^2_{e_3}) \over 3}.
\eeq
Note that $\varphi$ and $\varphi^{\prime}$ together break all of the
SM gauge symmetries, hence, they can affect thermalization of inflaton
decay products~\cite{AVERDI2}.

These two directions independently oscillate with amplitudes
$\varphi_0$ and $\varphi^{\prime}_0$, respectively, and frequencies
$m_{\varphi}$ and $m_{\varphi^{\prime}}$ respectively.

There are four light scalar degrees of freedom which correspond to
$\varphi$ and $\varphi^{\prime}$. The total number of scalars is
$N_{total} = 28$, with $18$ contained in the ${u_1d_1d_2}$ monomial
and $10$ in the ${L_1L_2e_3}$ monomial. Since the SM gauge group is
completely broken, we have $N_{broken} = 12$. Then
Eq.~(\ref{lightnum}) results in $N_{light} = 28 - 24 = 4$
.

This implies that $\varphi$ and $\varphi^{\prime}$ are the only light
fields around. Their mass eigenstates and eigenvalues, given by
Eqs.~(\ref{flat1}) and~(\ref{mass1}), respectively, are constant in
time and therefore not affected by their oscillations. Therefore, the
condition in Eq.~(\ref{nonad}) will not be satisfied. There are time
variations in the mass eigenstates and eigenvalues of the heavy
fields. However, as discussed in the previous section, so long as
$\varphi$ and $\varphi^{\prime}$ have rotational motion, resonant
particle production will be negligible~\cite{ACS,MP}.

There are many examples where the fields in the two monomials can have
Yukawa couplings to each other. In these cases simultaneous
$F$-flatness of the two directions is not guaranteed in general. For
an example, let us consider the two directions represented by $QQQL$
and $LLddd$ monomials. Note that $Q$ and $d$ can be coupled to each
other through quark Yukawas. If this is the case, then the two
directions cannot develop large VEVs simultaneously.

However, Yukawas couple fields with the same {\it color} index to each
other. One can then choose {\it color} indices such that $Q$ and $d$
from the same {\it family} have different {\it colors}.  This implies
that there is a subset of directions which are simultaneously $D$- and
$F$-flat, and hence, can develop large VEVs independently. The
situation will be similar for the directions represented by $udd$ and
$QLd$ monomials~\footnote{The second version of Ref.~\cite{OP} also
cites other examples of two independent flat directions such as $LLe$
and $QLd$. However, simultaneous $F$-flatness of these two directions
requires that the $L$ in $QLd$ must belong to the same {\it family} as
one of the two $L$'s in $LLe$.  Therefore one cannot have two
independent directions parameterized as $L = L = e = \varphi_0$ and $Q
= L = d =\varphi^{\prime}_0$ (up to two independent phases). It is
possible that $Q$, $L$, $d$, and $e$ all have a non-zero VEV, but the
situation will actually be similar to that of multiple flat
directions, which will be discussed later. This is also the case for
two directions represented by $LLe$ and $LLddd$ monomials.}.

In the above mentioned cases, one finds $N_{light} > 4$, which
implies that there are light degrees of freedom besides the two
oscillating flat directions.  This however is not a sufficient
condition to have preheating. To elucidate this point, let us
consider a toy model with a $U(1)$ gauge group and four scalar
fields $\phi_1,~\phi^{\prime}_1,~\phi_2,~\phi^{\prime}_2$ with
respective charges $+1,+1,-1,-1$. The corresponding $D$-term is
\beq \label{dterm2} V_D \sim g^2 \left({\vert \phi_1 \vert}^2 -
{\vert \phi_2 \vert}^2 + {\vert \phi^{\prime}_1 \vert}^2 - {\vert
\phi^{\prime}_2\vert}^2\right)^2. \eeq
After the following field redefinition
\begin{eqnarray} \label{redef}
\varphi = {(\phi_1 + \phi_2) \over \sqrt{2}} & , & \varphi^{\prime}
= {(\phi^{\prime}_1 + \phi^{\prime}_2) \over
\sqrt{2}}\, \nonumber \\
\chi = {(\phi_1 - \phi_2) \over \sqrt{2}} & , & \chi^{\prime} =
{(\phi^{\prime}_1 - \phi^{\prime}_2) \over \sqrt{2}}\, ,
\end{eqnarray}
we have~\footnote{A general $D$-term has the same structure, see
Eq.~(\ref{dterm}). It is therefore straightforward to generalize
this discussion made for a $U(1)$ gauge group.}
\beq \label{dterm3} V_D \sim g^2 \left(\varphi^{\prime}_R
\chi^{\prime}_R + \varphi^{\prime}_I \chi^{\prime}_I + \varphi_R
\chi_R + \varphi_I \chi_I \right)^2\,. \eeq
where $R$ and $I$ denote the real and imaginary components of a
scalar field respectively. We consider flat directions with maximal
rotation~\footnote{Generalization to the case with general rotation
is straightforward.}:
\begin{eqnarray} \label{maxrot}
\varphi_R = \varphi_0 {\rm cos} (m_{\varphi}t + \theta) & , & \varphi_I =
\varphi_0 {\rm sin}(m_{\varphi}t + \theta) \, \nonumber \\
\varphi^{\prime}_R = \varphi^{\prime}_0 {\rm cos}(m_{\varphi^{\prime}}t)
& , & \varphi^{\prime}_I = \varphi^{\prime}_0 {\rm sin}
(m_{\varphi^{\prime}}t) \, ,
\end{eqnarray}
where $\theta$ is a constant phase.
There is a total of $8$ scalar degrees of freedom. Four of
them $\varphi_R,~\varphi_I,~\varphi^{\prime}_R,~\varphi^{\prime}_I$
represent the two complex flat directions. The mass matrix for the
other four degrees of freedom can be diagonalized to obtain
instantaneous mass eigenstates. It is readily seen from
Eq.~(\ref{dterm3}) that one degree of freedom
\begin{eqnarray} \label{chi1}
\chi_1 (t) &=&(\varphi_0^2 + {\varphi^{\prime}_0}^{2})^{-1/2}
\Big[\varphi^{\prime}_0 {\rm cos} (m_{\varphi^{\prime}}t)
\chi^{\prime}_R + \varphi^{\prime}_0 {\rm sin}
(m_{\varphi^{\prime}}t)
\chi^{\prime}_I \, \nonumber \\
&+&\varphi_0 {\rm cos}(m_{\varphi}t + \theta)
\chi_R + \varphi_0 {\rm sin}(m_{\varphi}t + \theta)\chi_I \Big]\, ,
\end{eqnarray}
has a mass $\sim g (\varphi_0^2 + {\varphi^{\prime}_0}^2)^{1/2}$ and
is superheavy. Spontaneous breakdown of the $U(1)$ by the VEV of two
flat directions yields a Goldstone boson
\begin{eqnarray} \label{chi2}
\chi_2 (t) &=&(\varphi_0^2 + {\varphi^{\prime}_0}^{2})^{-1/2} \Big[-
\varphi^{\prime}_0 {\rm sin} (m_{\varphi^{\prime}}t) \chi^{\prime}_R
+ \varphi^{\prime}_0 {\rm cos} (m_{\varphi^{\prime}}t)
\chi^{\prime}_I \, \nonumber \\
&-&\varphi_0 {\rm sin}(m_{\varphi}t + \theta)
\chi_R + \varphi_0 {\rm cos}(m_{\varphi}t + \theta)\chi_I \Big]\, ,
\end{eqnarray}
which is eaten up via the Higgs mechanism (hence removed from the
spectrum). Note that in the special cases where $\varphi_0 = 0$ or
$\varphi^{\prime}_0 = 0$, the expression in Eq.~(\ref{chi2}) is
reduced to that for a single flat direction, given in the second
line of Eq.~(\ref{eigst}).

Finally, there are two physical light degrees of freedom
\begin{eqnarray} \label{chi3}
\chi_3 (t) = (\varphi_0^2 + {\varphi^{\prime}_0}^{2})^{-1} \Big(a_1
\chi^{\prime}_R + a_2 \chi^{\prime}_I +  a_3 \chi_R + a_4 \chi_I
\Big) \, , \nonumber \\
\,
\end{eqnarray}
and
\begin{eqnarray} \label{chi4}
\chi_4 (t) = (\varphi_0^2 + {\varphi^{\prime}_0}^{2})^{-1} \Big(b_1
\chi^{\prime}_R + b_2 \chi^{\prime}_I + b_3 \chi_R + b_4 \chi_4
\Big) \, , \nonumber \\
\,
\end{eqnarray}
where
\begin{eqnarray} \label{a}
a_1 &=&- {\varphi_0}^2 + \varphi_0 \varphi^{\prime}_0 ~ {\rm cos} (\Delta m t + \theta) \, \nonumber \\
a_2 &=&- \varphi_0 \varphi^{\prime}_0 ~ {\rm sin} (\Delta m t + \theta) \,  \nonumber \\
a_3 &=& - {\varphi^{\prime}_0}^2 + \varphi_0 \varphi^{\prime}_0 ~ {\rm cos} (\Delta m t + \theta) \, \nonumber \\
a_4 &=&\varphi_0 \varphi^{\prime}_0 ~ {\rm sin} (\Delta m t +
\theta) \, ,
\end{eqnarray}
and
\begin{eqnarray} \label{b}
b_1 &=&- \varphi_0 \varphi^{\prime}_0 ~ {\rm sin} (\Delta m t + \theta) , \nonumber \\
b_2 &=&{\varphi_0}^2 - \varphi_0 \varphi^{\prime}_0 ~ {\rm cos} (\Delta m t + \theta) \, \nonumber \\
b_3 &=&\varphi_0 \varphi^{\prime}_0 ~ {\rm sin} (\Delta m t + \theta) \, \nonumber \\
b_4 &=& {\varphi^{\prime}_0}^2 - \varphi_0 \varphi^{\prime}_0 ~ {\rm
cos} (\Delta m t + \theta) \, .
\end{eqnarray}
These two eigenstates have a soft supersymmetry breaking mass
$m_{\chi} \sim {\cal O}({\rm TeV})$. Here $\Delta m \equiv
m_{\varphi} - m_{\varphi^{\prime}}$ is the mass difference of the
two flat directions.

Time variation in $\chi_3(t),~\chi_4(t)$ may lead to resonant
particle production. Note however that this can happen only if
$m_{\varphi} \neq m_{\varphi^{\prime}}$. Indeed it can be seen from
Eqs.~(\ref{chi3},\ref{chi4}) that there will be no time-dependent
terms in $\chi_3(t),~\chi_4(t)$ if the flat directions have the same
mass. This can be qualitatively understood as following. The two
rotating flat directions can be visualized as two vectors which
rotate in the field space. So long as the flat directions have the
same mass, the two vectors rotate with the same frequency, and so
does the resultant vector from their addition. Therefore the
situation is effectively reduced to that of one rotating flat
direction. In this case, as explained in the previous section, all
time variation is in the sector of heavy fields. Neither the mass
eigenstates nor the mass eigenvalues of the physical light fields
are time-dependent. In consequence, there will be no resonant
particle production.

Having $\Delta m \neq 0$ is necessary but not sufficient for
significant particle production. To demonstrate this let us consider
two flat directions with hierarchically different VEVs. Without loss
of generality, we assume that $\varphi^{\prime}_0 \ll \varphi_0$.
Then, to the leading order in $\varphi^{\prime}_0/\varphi_0$, we
have
\begin{eqnarray} \label{3n}
\chi_3 (t) = - \chi^{\prime}_R  &+& \Big({\varphi^{\prime}_0 \over
\varphi_0}\Big) \, \nonumber \\
\Big[{\rm cos} (\Delta m t + \theta) (\chi^{\prime}_R + \chi_R) &+&
{\rm sin} (\Delta m t + \theta) (- \chi^{\prime}_I + \chi_I)
\Big] \, , \nonumber \\
\,
\end{eqnarray}
and
\begin{eqnarray} \label{4n}
\chi_4 (t) = \chi^{\prime}_I &+& \Big({\varphi^{\prime}_0 \over
\varphi_0}\Big) \, \nonumber \\
\Big[{\rm sin} (\Delta m t + \theta) (-\chi^{\prime}_R + \chi_R) &-&
{\rm cos} (\Delta m t + \theta) (\chi^{\prime}_I + \chi_I) \Big] \, . \nonumber \\
\,
\end{eqnarray}
The important point is that $\varphi^{\prime}_0/\varphi_0$ appears
in front of the time-dependent terms in
Eqs.~(\ref{3n},\ref{4n})~\footnote{In the opposite case when
$\varphi_0 \ll \varphi^{\prime}_0$, the expressions for $\chi_3 (t)$
and $\chi_4 (t)$ are obtained by changing $\varphi \leftrightarrow
\varphi^{\prime}$, $\chi_{R,I} \leftrightarrow \chi^{\prime}_{R,I}$,
$(\Delta m t + \theta) \rightarrow -(\Delta m t + \theta)$ in
Eqs.~(\ref{3n},\ref{4n}).  The time-dependent terms are in this case
suppressed by $\varphi_0/\varphi^{\prime}_0 \ll 1$.}. Temporal
derivatives of these terms which show up in the equation of motion
of $\chi_3(t),~\chi_4(t)$ modes will therefore be suppressed by this
factor. It follows from the analysis of~\cite{OP} that modes with a
physical momentum $k \leq \Delta k$ will undergo resonance where
\beq \label{reswid} \Delta k \simeq \sqrt{\Big({\varphi^{\prime}_0
\over \varphi_0}\Big) (m_{\varphi} - m_{\varphi^{\prime}})^2 -
m^2_{\chi}}. \eeq
Had $\chi_3(t),~\chi_4(t)$ been massless (i.e. $m_{\chi}=0$), the
width of the resonance band would have shrunk $\propto
(\varphi^{\prime}_0/\varphi_0)^{1/2}$. Modes within the resonance
band were exponentially amplified, though the required time scale
would increase by a logarithmic factor ${\rm
log}(\varphi_0/\varphi^{\prime}_0)$ due to phase space suppression.
However supersymmetry breaking inevitably results in a soft mass
$m_{\chi} \sim {\cal O}({\rm TeV})$. Hence the resonance band
completely disappears when~\footnote{This point is ignored in the
third version of~\cite{OP}, see its added note.}
\beq \label{resvan} {\varphi_0 \over \varphi^{\prime}_0} \gs
\Big({m_{\varphi} - m_{\varphi^{\prime}} \over m_{\chi}}\Big)^2.
\eeq
Since $m_{\chi},~m_{\varphi},~m_{\varphi^{\prime}}$ are all set by
supersymmetry breaking, even a small hierarchy between
$\varphi^{\prime}_0$ and $\varphi_0$ will be sufficient to kill off
the resonance band. Typically there will be no non-perturbative
particle production if $\varphi_0 \geq 10
\varphi^{\prime}_0$~\footnote{Note that Hubble expansion redshifts
the VEV of two flat directions the same. Hence Eq.~(\ref{reswid})
determines the width of resonance band throughout the expansion of
Universe.}.

We remind that the VEV of a flat direction at the onset of its
oscillations is set by the higher-order superpotential term of the
form $\varphi^n/M^{n-3}$ (induced by physics at a high scale $M$)
which lifts it: $\varphi_0 \sim (m_{\varphi}
M^{n-3})^{1/n-2}$~\cite{DRT}. Directions which are represented by
different monomials usually are lifted by terms with different
values of $n$, and hence have different VEVs. For example, ${QLd}$
and ${QQQL}$ are lifted by $n=4$ terms, while ${udd}$ and ${LLddd}$
survives until $n=6$ and $n=7$ respectively~\cite{GKM}. Consider any
two such flat directions $\varphi^{\prime}$ and $\varphi$ which are
lifted by superpotential terms of order $m$ and $n$ respectively ($3
< m < n$). For $M = M_{\rm P}$ or $M = M_{\rm GUT}$, we indeed have
$\varphi_0 \geq 10 \varphi^{\prime}_0$ for all cases with $m \neq
n$. Then, as we explained above, there will be no preheating at all.

One comment is in order before closing this subsection. As we
discussed in section 1, the masses induced by the VEV of flat
directions are supersymmetry conserving. This implies that the
spectrum of fermions is the same as that of bosons except for the
soft supersymmetry breaking effects. In particular, the eigenstates
of fermionic partners of $\chi_3(t)$ and $\chi_4(t)$, denoted by
${\tilde \chi}_3(t)$ and ${\tilde \chi}_4(t)$ respectively, have the
same time dependence as in Eqs.(\ref{3n},\ref{4n})~\footnote{Note
that we are only interested in the light fields. Hence we do not
consider gauge and gaugino fields as they have a mass $\sim g
\varphi_0$ and are superheavy.} . Hence, similar to
Eq.~(\ref{reswid}), corresponding modes with a physical momentum $k
\leq {\Delta k}_{ferm}$ have resonance instability where
\beq \label{fermres} {\Delta k}_{ferm} \simeq
\Big({\varphi^{\prime}_0 \over \varphi_0}\Big)^{1/2} \vert
m_{\varphi} - m_{\varphi^{\prime}} \vert. \eeq
The important difference is that fermionic modes have no
supersymmetry breaking mass $m_{\chi}$. Thus the resonance band
shrinks as a result of a hierarchy $\varphi^{\prime}_0 \ll
\varphi_0$ but, unlike the bosonic case, does not completely
disappear. Therefore one might think that flat directions will
quickly decay via non-perturbative production of fermions.

Let us estimate the rate for production of fermions
The key point is that, due to Pauli blocking, the occupation number
of fermions cannot exceed $1$. An occupation number $\sim 1$ is
achieved within a time $\delta t \sim {\Delta
k}^{-1}_{ferm}$~\cite{OP}. The energy density in the produced
particles is $\rho_{ferm} \sim {\Delta k}^4_{ferm}$. Hubble
expansion and rescatterings deplete the resonance band, after which
production of fermions within the band will resume.
Assuming that this happens immediately, an amount ${\Delta
k}^4_{ferm}$ of energy density is transferred to fermions within
each time interval $\sim {\Delta k}^{-1}_{ferm}$. Then the energy
density in fermions increases by $\sim ({\Delta k}^5_{ferm}/H)$ in
one Hubble time. Particle production becomes efficient when this is
comparable to the energy density in flat directions~\footnote{For
hierarchical VEVs, i.e. $\varphi_0 \gg \varphi^{\prime}_0$, the
energy density is mainly in $\varphi$.}. In a Universe dominated by
relativistic particles (from inflaton decay), the latter changes
$\propto H^{3/2}$. With the help of Eq.~(\ref{fermres}), it turns
out that production of fermions will be efficient when $H \ll
m_{\varphi} (m_{\varphi}/\varphi_0)^{4/5}$.
%
%
%
%
This is sufficiently late in order for the flat directions to affect
thermal history of the Universe~\cite{AVERDI2}.


\subsection{Multiple directions represented by a polynomial}

When the family indices of lepton and quark multiplets are taken into
account, we will have a gauge-invariant polynomial which represent
multiple flat directions~\footnote{In this sense the ${H_u H_d}$
monomial is very special, since it is the only one which has no family
indices.}. For example, let us consider the ${ H_u L_i}$ polynomial,
where $1 \leq i \leq 3$ (for a detailed discussion, see~\cite{JKM}).

Assuming $D$-flatness, one can always go to a basis where $H^1_u =
L^2_1 = L^2_2 = L^2_3 = 0$ (superscripts denote weak isospin
components), and (up to an overall phase)
\begin{eqnarray} \label{hl2}
& H^2_u = {\varphi_0 \over \sqrt{2}}, \, \nonumber \\
& \, \nonumber \\
& L^1_1 = \frac{a \varphi_0}{\sqrt{2}}, ~ L^1_2 =
\frac{b \varphi_0}{ \sqrt{2}},  ~
L^1_3 = \frac{\sqrt{1 - \vert a
\vert^2 - \vert b \vert^2} \varphi_0}{\sqrt{2}}, \,
\end{eqnarray}
where $\vert a \vert^2 + \vert b \vert^2 \leq 1$.

A flat direction is then given by
\beq \label{hlflat}
\varphi = {(H^2_u + a^{\ast} L^1_1 + b^{\ast} L^1_2 +
\sqrt{1 - \vert a \vert^2 - \vert b \vert^2} L^1_3) \over \sqrt{2}},
\eeq
and its mass is:
\beq \label{hlmass}
m^2_{\varphi} = {m^2_{H_u} + \vert a \vert^2 m^2_{L_1} + \vert b \vert^2
m^2_{L_2} + (1 - \vert a \vert^2 - \vert b \vert^2) m^2_{L_3} \over 2}.
\eeq
Now the flat manifold has complex dimension $3$. Note that $N_{total}
= 16$ and $N_{broken} = 3$, and hence Eq.~(\ref{lightnum}) results in
$N_{light} = 10$. This implies that there are four light degrees of
freedom besides those which parameterize the flat manifold.

The trajectory of the flat direction VEV is determined by its initial
value, the soft breaking terms and higher order superpotential terms
which lift the flatness. The flat direction motion is confined to a
complex plane when $a,~b$ are constant in time (in particular for
$a=b=0$, $a=0$ and $\vert b \vert =1$, or $\vert a \vert =1$ and
$b=0$). However, for a general motion on the flat manifold, $a$ and
$b$ can assume any values in the allowed ranges, and hence are time
dependent. The phase motion of the fields was already recognized in
Ref.~\cite{JKM}.

However, note that just time variation of mass eigenstates is not a
sufficient condition for a quick non-perturbative decay of the flat
direction. A closer analysis suggests that the manifold spanning the
flat direction is curved and the kinetic terms are
non-minimal~\cite{JKM}.  It was already noticed in Ref.~\cite{JKM},
that a small perturbation of the flat direction trajectory will be
highly non-trivial, the trajectories have instability which leads to
possibly a chaotic motion. Particle production due to a chaotic motion
is much more involved than that for an oscillatory motion and, a
detailed investigation is needed to clarify this situation.

Further note, that the other $4$ light scalars (besides $6$, which
parameterize the flat manifold) are expected to have
(self-)interactions with gauge strengths (from the $D$-terms), as they
do not correspond to genuinely flat directions, see
Eq.~(\ref{hl2},\ref{hlflat}).  Such interactions can strongly suppress
non-perturbative particle production~\cite{AC,PR}.

An important point is to note that the flat manifold does not remain
the same for all field values. As mentioned, higher-order
superpotential terms of the form, $\varphi^n/M^{n-3}$ (with $n > 3$),
lift the flatness for large field values. If one writes all higher
order terms which respect the SM gauge symmetries, then none of the
MSSM flat directions survive beyond $n=9$~\cite{GKM}.  A higher order
term $\varphi^n/M^{n-3}$ becomes important at a field value $\vert
\varphi \vert \sim (m_{\varphi} M^{n-3})^{1/n-2}$~\cite{DRT,GKM}.
Therefore, terms with different $n$ take action at different VEVs and
lifting of the flatness happens at stages. In some physically
interesting cases the flat manifold has complex dimension ``one'' at
very large field values, which effectively leaves us with a {\it
single} flat direction.  This happens for the $LLe$, $LLddd$ and
$QuQue$ polynomials (when all families are taken into account).  The
flat manifolds corresponding to these polynomials have complex
dimension $3$, $3$, and $16$ respectively. However beyond $n=4$,
equivalently at field values larger than $(m_{\varphi} M)^{1/2}$, they
all become effectively of complex dimension one~\cite{GKM}. These
remaining {\it single} flat directions are eventually lifted with
$n=6$, $n=7$, and $n=9$ terms respectively~\cite{GKM}. Then, starting
at large VEVs, the situation will be similar to that for a single flat
direction discussed in the previous section, and hence there will be
no preheating.

Finally, as a side comment, we emphasize that the flat directions can
``terminate'' an early stage of non-perturbative inflaton decay via
preheating, regardless of their fate~\cite{AVERDI3}. They can induce
large masses to the inflaton decay products at time scales long before
the flat direction oscillations start. Hence, contrary to the standard
lore~\cite{LINDE}, the inflaton can decay perturbatively even if it
has large couplings to the matter fields.

Within MSSM the only renormalizable interactions of a {\it gauge
singlet} inflaton, ${ \Phi}$, with matter can happen through
superpotential terms of the form $h { \Phi H_u L}$ or $h { \Phi H_u
H_d}$~\cite{AVERDI3}.  Rest of the superpotential couplings to matter
fields are through non-renormalizable terms. The couplings of a
singlet to the matter fields through derivative terms (dictated by
K\"ahler potential) will also be non-renormalizable.

In models of large field inflation, a typical VEV of the inflaton is
$\langle \Phi \rangle \sim {\cal O}(M_{\rm P})$, during and just after
inflation. Then any renormalizable couplings of the inflaton induces a
large SUSY conserving mass, $h M_{\rm P}$, for ${ H_u}$ and ${ L}$, or
${ H_u}$ and ${ H_d}$ multiplets. This implies that neither ${ H_u L}$
nor ${ H_u H_d}$ directions can develop a large VEV during
inflation. Hence neither of them are relevant in the process of
thermalization, unless the inflaton is very weakly
(i.e. gravitationally) coupled to all fields.

\section{Conclusions}

To conclude, the particle production via preheating does not happen
for a single flat direction, it is also not likely for multiple flat
directions. In many cases which are of physical interest, after
taking into account of the $F$-term constraints and $F$-term
lifting, the multiple flat direction case becomes effectively the
same as that of a single flat direction, or collection of
independent {\it single} flat directions. In either cases preheating
is not relevant. In particular, the flat direction which has the
largest VEV survives long. The longevity of the flat direction then
ensures that thermalization of the inflaton decay products is
considerably modified~\cite{AVERDI1,AVERDI2}.

The long lifetime of MSSM flat directions leads to late thermalization
of inflaton decay products and a low reheat temperature. This gives
rise to a natural solution to the thermal overproduction of
gravitinos, and other dangerous
relics~\cite{AVERDI1,AVERDI2}. Regardless of their longevity, flat
directions can also lead to a perturbative decay of a gauge singlet
inflaton, even if it has large couplings to matter
fields~\cite{AVERDI3}. To summarize, the MSSM flat directions in
general play crucial role in reheating and thermalization.

\section{Acknowledgments}

The authors wish to thank Asko Jokinen, Keith Olive and Marco Peloso
for useful discussions. The work of R.A. is supported by the National
Sciences and Engineering Research Council of Canada (NSERC).


\end{document}